\documentclass{nature}
\usepackage{color}
\usepackage{dcolumn}
\usepackage{amsmath}
\usepackage{units}
\usepackage{graphicx}
\makeatletter
\let\saved@includegraphics\includegraphics
\AtBeginDocument{\let\includegraphics\saved@includegraphics}
\renewenvironment*{figure}{\@float{figure}}{\end@float}
\makeatother

\usepackage{caption}

\begin{document}

\title{Lifted electron pocket and reversed orbital occupancy imbalance in FeSe}

\author{S. S. Huh$^{1,2}$, J. J. Seo$^{3}$, B. S. Kim$^{1,2}$, S. H. Cho$^{1,2}$, J. K. Jung$^{1,2}$, S. Kim$^{4}$, Y. Y. Koh$^{5}$, C. I. Kwon$^{5,6}$, Jun Sung Kim$^{5,6}$, W. S. Kyung$^{7}$, J. D. Denlinger$^7$, Y. H. Kim$^8$, B. N. Chae$^8$, N. D. Kim$^8$, Y. K. Kim$^{4,*}$,  C. Kim$^{1,2,*}$}


\maketitle

\begin{affiliations}
\item Department of Physics and Astronomy, Seoul National University (SNU), Seoul 08826, Republic of Korea
\item Center for Correlated Electron Systems, Institute for Basic Science (IBS), Seoul 08826, Republic of Korea
\item Institute of Physics and Applied Physics, Yonsei University, Seoul 03722, Korea
\item Department of Physics, Korea Advanced Institute of Science and Technology, Daejeon 34141, Republic of Korea
\item Department of Physics, Pohang University of Science and Technology, Pohang 37673, Republic of Korea
\item Center for Artificial Low Dimensional Electronic Systems, Institute for Basic Science (IBS), Pohang 37673, Republic of Korea
\item Advanced Light Source, Lawrence Berkeley National Laboratory, Berkeley, CA 94720, USA
\item Pohang Accelerator Laboratory, Pohang University of Science and Technology, Pohang 37673, Republic of Korea
\end{affiliations}

\noindent

\vspace{10 pt}
\vspace{10 pt}

{\bf The FeSe nematic phase has been the focus of recent research on iron based superconductors (IBSs) due to its unique properties. A number of electronic structure studies were performed to find the origin of the phase. However, such attempts came out with conflicting results and caused additional controversies. Here, we report results from angle resolved photoemission and X-ray absorption spectroscopy studies on FeSe with detwinning by a piezo stack. We have fully resolved band dispersions with orbital characters near the Brillouin zone corner which reveals absence of a Fermi pocket at the Y point in the 1Fe Brillouin zone. In addition, the occupation imbalance between $d_{xz}$ and $d_{yz}$ orbitals is found to be opposite to that of iron pnictides, which is consistent with the identified band characters. These results settle down controversial issues in the FeSe nematic phase and shed light on the origin of nematic phases in IBSs.}

Nematic phase is a state with a broken rotational symmetry but with an intact translational symmetry. It has attracted renewed attention with a notion that it may provide an important clue to the mystery of unconventional superconductivity\cite{nem}. Its region almost coincides with that of the superconducting region in the phase diagram for both cuprates \cite{cup} and iron based superconductors (IBS) \cite{ibsnem}, implying its possible connection to the superconductivity. Finding the origin of nematic phases thus has been considered to be one of the most important goals in the research on unconventional superconductivity, especially for IBSs where studies on nematic phases were first initiated.

Recently, the nematic phase in FeSe has attracted attention due to its distinct properties from those of pnictide nematic phases. The most peculiar aspect is the absence of long range magnetic order which always coexists with orbital order in pnictide nematic phases \cite{nomag1, nomag2}. Furthermore, it was revealed that resistivity anisotropy for FeSe has the opposite sign compared to that of iron pnictides\cite{fsres}; resistivity along the longer $a$-axis is smaller than that along the shorter $b$-axis for iron pnictides while it is the other way around for FeSe. These two observations prompted the conjecture that the nematic phase in FeSe may be different from that of iron pnictides. Therefore, understanding the FeSe nematic phase may provide insight on universal understanding of the nematic phase in IBS. More importantly, considering the recent discovery of orbital selective correlation and superconductivity in FeSe in STM studies\cite{stm1,stm2}, superconductivity mechanism could be addressed by understanding the origin of the nematic phase because the nematic phase is widely believed to induce orbital selectivity.

In this regard, a number of angle resolved photoemission spectroscopy (ARPES) experiments have been performed to investigate the electronic structure of FeSe\cite{ferro,dwave,uni,sr}. However, interpretations from different experiments vary and caused more controversies on the origin of the nematic phase in FeSe. Various conflicting scenarios were proposed as the origin of the nematic phase based on ARPES results such as simple ferro-orbital ordering\cite{ferro}, d-wave orbital order\cite{dwave}, unidirectional nematic bond order\cite{uni} and reversal sign ordering\cite{sr}. Probable cause of such controversies is the lack of full and accurate electronic structure information which may be obtained only from fully detwinned single crystals. Therefore, the issue may be resolved only with full electronic structure information. Only then, the origin of the nematic phase in FeSe may be addressed.

Here, we present results of electronic structure studies by AREPS and X-ray linear dichroism (XLD) on fully detwinned FeSe in its nematic phase. Full detwinning was obtained by using a piezo stack based strain device, the first attempt to use such device in ARPES and XLD experiments (See Figs. 1A and 1B). Difficulty in using mechanical strain methods such as uncertainty of strain direction from 'accident detwin' by anisotropic shrink of epoxy \cite{det} has been overcome by using piezo bias on/off which provides strain on/off. As a result, we could investigate detailed band dispersions with full orbital characters (with ARPES) as well as the orbital occupancy imbalance between $d_{xz}$ and $d_{yz}$ orbitals (by XLD). Our ARPES results show that there are orbital dependent band shift and hybridization that lead to only one Fermi surface pocket in 1Fe BZ scheme, which may explain why there is no long-range magnetism. Furthermore, XLD reveals an unexpected reversed occupation imbalance between $d_{xz}$ and $d_{yz}$ orbital ($n_{xz}$ $<$ $n_{yz}$) which naturally explains the opposite resistivity anisotropy. These results resolve controversial issues on the nematic phase of FeSe and may shed light on the driving instability of nematic phases in IBS.

\noindent
\newline
{\bf Results}
\newline
{\bf Electronic structures of twinned and detwinned FeSe.} The key issue in the study of the FeSe electronic structure is if, in addition to an elliptical pocket, there is another large pocket which mainly consists of the $d_{xy}$ orbital around the Brillouin zone (BZ) corner. Therefore, we first focus on the overall Fermi surface topology. Fermi surface maps from twin and single domains (Figure 1c) indeed show clear differences, and may provide answers to the issue. For the twin domain case (left), two perpendicularly crossing elliptical pockets are observed at zone corner X/Y points. However, Fermi surfaces from a single domain sample (right) consists only of a single elliptical pocket at each zone corner. Note that, if there is a large pocket of $d_{xy}$ orbital character, it should be visible in our data as the experimental geometry allows the transition from the $d_{xy}$ initial state. Based on this experimental observation as well as the full band dispersion characterization along with its temperature dependence (discussed below), we conclude that there is only one elliptical pocket at zone corners as illustrated in Figure 1d. It immediately implies that, one of two pockets in the normal state should disappear across the nematic phase transition. That is, the pocket at the Y point in 1Fe BZ scheme shown in Figure 1e should disappear while the pocket at the X point remains.

\noindent
\newline
{\bf Dispersions and orbital characters of bands.} To proceed with the discussion on how and why the pocket at the Y point in 1Fe BZ disappears, band dispersions as well as orbital characters should be fully identified first. Figures 2a and 2b show overall band dispersions along the $\Gamma$-X and $\Gamma$-Y directions from a single domain sample in the normal and nematic states. Starting from isotropic normal state band dispersions, highly anisotropic band dispersions develop in the nematic phase. Especially, the band dispersions around the zone corner are dramatically renormalized and become those of two merged Dirac cones (possibly with small gaps at band crossing points) (see Figure 2b). The $\Gamma$-X ($k_{x}$-direction) data shows a sizable electron band and two split hole bands, while a tiny electron band near $E_{F}$ and two closely located hole bands at a higher binding energy are seen in the cut along the $\Gamma$-Y ($k_{y}$-direction). These observed dispersions are consistent with previous results \cite{det}. In order to understand the band structure more concretely, polarization dependent experiment is required to provide exact orbital characterization. The results of orbital characterization of bands are shown in Figures. 2c-2j. In particular, orbital characters of bands near the zone corner are newly interpreted. Along the $k_{x}$-direction, the large electron band consists mainly of $d_{xy}$ orbital while the two upper and lower hole bands have $d_{yz}$ and $d_{xy}$ characters. Along the ($k_{y}$-direction, the tiny electron band consists $d_{yz}$ orbital and hole bands have $d_{xz}$ and $d_{xy}$ characters (see Figures 2e and 2i). As for the zone center, the polarization dependence yields orbital characters of the hole bands that are consistent with previous results (See Figures 2f and 2j) \cite{sr}.

The resulting full orbital characters of the bands are schematically summarized in Figures 2k and 2l. The result reveals two unique features of the FeSe electronic structure in the nematic phase. First of all, there is a reversal behavior in the relative energy positions of $d_{xz}$ and $d_{yz}$ hole bands. The $d_{xz}$ hole band at Y point is located below the $d_{yz}$ hole band at the X point (zone corners) while the $d_{xz}$ hole band is placed above the $d_{yz}$ band at the zone center. The other and more important feature is the reduced number of electron bands at X and Y points. In the normal state, there are two electron bands at both X and Y points - the common $d_{xy}$ electron band and $d_{xz}$ ($d_{yz}$) electron band for the X (Y) point. On the other hand, only one electron band exists at each point in the nematic phase: $d_{xy}$ ($d_{yz}$) at the X (Y) point. Other two electron bands - $d_{xz}$ and $d_{xy}$ at X and Y points, respectively - are not observed in any experimental geometry, i.e. polarization. Therefore, it is reasonable to speculate that those two bands do not exist below $E_{F}$, and that they should be pushed above $E_{F}$ across the nematic phase transition.

\noindent
\newline
{\bf Temperature evolution of electronic structure.} The temperature evolution of electronic structure across the nematic phase transition clearly shows that what we speculated is indeed the case. Figure 3a shows temperature dependence of band dispersion along $k_{x}$-direction at the Y point, with overlaid momentum distribution curves taken at 5meV above $E_{F}$. In the normal state, $d_{xz}$ electron band crosses $E_{F}$ and peak positions in the curve have finite momentum values (indicated with red arrows). As the temperature decreases, peak positions for the $d_{xz}$ electron band shift towards the Y point. This indicates that the $d_{xz}$ electron band shifts upward and is finally pushed above $E_{F}$. Meanwhile, the $d_{yz}$ ($d_{xz}$) hole band around the X (Y) point shifts upward (downward) as the temperature decreases, as seen in temperature dependent spectra at X and Y points (Figures. 3b and 3c). The splitting at a low temperature is about 60 meV and remains finite above the nematic phase transition temperature, possibly due to the applied strain (Figure 3c). It is noteworthy that the two hole bands do not cross $E_{F}$ and remain under $E_{F}$ after the shift.

Now we are ready to discuss how the electron pocket at Y disappears in 1Fe BZ, that is, how the number of electron bands is reduced across the nematic phase transition. We argue that the observed electronic structure evolution can be explained within the picture of orbital dependent band shift (or splitting) and hybridization. Let us look at the case for the zone corner in 1Fe BZ scheme. Starting from symmetric bands in the normal state (Figure 3d left), the $d_{yz}$ ($d_{xz}$) band shifts upward (downward) as the temperature goes down below $T_{S}$ (Figure 3d middle), which we call ‘nematic band shift’. Then, the orbital dependent hybridization comes in. For the $d_{yz}$ band case, there is only weak mixing with $d_{xy}$ and thus the overall dispersion remains intact. The $d_{xz}$ band, on the other hand, strongly hybridizes with $d_{xy}$ band, pushing the hole band down below $E_{F}$ while lifting both $d_{xz}$ and $d_{xy}$ electron bands above $E_{F}$ (Figure 3d right). This, results in vanishing electron pocket at Y in the 1Fe BZ scheme.

\noindent
\newline
{\bf Reversed orbital occupancy imbalance.} A surprising implication of the above interpretation is that $d_{xz}$ orbital should be less occupied than $d_{yz}$, opposite to the iron pnictide case \cite{yk} and also to the prediction of ferro-orbital order scenario \cite{ccl}. At the zone center, the energy position of the $d_{xz}$ band is higher than that of $d_{yz}$, and thus $d_{xz}$ state is less occupied. Disappearance of the $d_{xz}$ electron band at the zone corner also leads to less occupied $d_{xz}$ orbital ($d_{xz}$ and $d_{yz}$ hole bands at the zone corner are irrelevant as both of them are fully occupied). In order to obtain direct information on such anomalous orbital occupancy, we performed XLD measurements on detwinned samples. XLD measurement, a local probe for orbital selective density of states, can provide direct proof of the imbalance in the orbital occupancy \cite{yk, ccc}.

Figure 4a shows the experimental geometry with two light polarizations, parallel and perpendicular to the strain direction. Fe $L$-edge absorption spectra from detwinned FeSe taken with the two light polarizations at 10 K are plotted in Figure 4b. With the given experimental geometry, the linear dichroism or XLD shown with the black solid line in Figure 4b should reflect the imbalance in $d_{xz}$ and $d_{yz}$ orbital occupancy. A complication is that not only the orbital occupation imbalance but also the orthorhombic structural distortion is known to contribute to the XLD signal \cite{ccc}. It was previously shown that the two contributions can be separated by considering their distinct behaviors in the temperature dependence \cite{yk}. A close inspection of the temperature dependent XLD data in Figure 4c reveals that XLD signal starts to appear below $T_{S}$ and monotonically increases as the temperature decreases down to 10 K. Such monotonic increase of XLD can be more clearly visualized by plotting the integrated XLD (see the figure caption for the definition) as a function of temperature (Figure 4d). If the XLD contains only the structure contribution, it should immediately saturate below $T_{S}$ as the structure contribution should follow the orthorhombicity of the crystal (see the overlaid diffraction data) \cite{xrd}. Therefore, the non saturating increase far below $T_{S}$ indicates that XLD does contain contribution from orbital imbalance. An important point to note is that the orbital contribution is positive. Positive XLD from orbital means that $d_{xz}$ orbital is less occupied ($n_{xz}$ $<$ $n_{yz}$), contrary to the case of simple ferro-orbital order scenario and to the iron pnictide case ($n_{xz}$ $>$ $n_{yz}$) (Figure 4e).

\noindent
\newline
{\bf Discussion}
\newline
\indent
As our interpretation of the electronic structure evolution across the nematic phase transition in FeSe is confirmed by observation of the reversed orbital occupancy imbalance, its implication may be discussed. First of all, the sign reversal in the hole band splitting and the reduced number of electron bands in our interpretation do not support the unidirectional nematic bond order scenario which requires absence of hole band splitting\cite{uni}. On the other hand, d-wave form splitting\cite{dwave} and sign reversal order\cite{sr} scenario are partially consistent with our result when the hole band splitting is considered. From these results, we learn that the evolution of electron band dispersions as well as the role of the $d_{xy}$ band are taken into account to obtain a fully consistent picture, that is to say, full understanding of the nematic phase.

With the concrete understanding of the electronic structure, the unique properties of FeSe nematic phase - absence of magnetism and opposite resistivity anisotropy - can also be understood. The absence of the magnetism can be explained for both weak and strong coupling pictures. In the weak coupling picture, a weak Fermi surface nesting condition stemming from opposite orbital characters of Fermi surface pockets could explain the absence of magnetism; the inter-orbital nesting between Fermi surfaces at the zone center (mostly $d_{xz}$ with small contribution from $d_{yz}$) and corner (mostly $d_{yz}$), which is considered to be the source of the magnetism in the weak coupling picture, is mostly suppressed due to the opposite orbital characters. In the strong coupling picture, the absence of the magnetism could be explained within an orbital weight redistribution scheme\cite{weak1,weak2}. A strong hybridization between $d_{xy}$ and $d_{xz}$ tends to open a gap near the $E_{F}$, resulting in suppression of the $d_{xy}$ and $d_{xz}$ orbital weight. Since the magnetic moment dominantly comes from the $d_{xy}$ orbital \cite{strong1,strong2}, the reduced $d_{xy}$ orbital weight should weaken the magnetic instability. Meanwhile, the opposite resistivity anisotropy can be easily explained within the observed reversed orbital occupancy imbalance if we simply follow the argument that orbital occupation imbalance characterizes the resistivity anisotropy \cite{resoo1,resoo2}.

Another important implication of our work is that it provides a new perspective on the origin of the nematic phase. It has been believed that the nematic band shift and the occupancy imbalance between $d_{xz}$ and $d_{yz}$ orbitals are equivalent. Therefore, they represent a single phenomenon of the ferro-orbital order stemming from symmetry breaking at the atomic level. Our observation of the reversed orbital occupancy imbalance, an occupancy imbalance opposite to the nematic band shift, clearly shows that they are not equivalent. Furthermore, their opposite behaviors strongly suggest that the nematic band shift is neither a manifestation of nor generated by the development of occupancy imbalance. The occupancy imbalance should rather be a by-product of the nematic band shift as well as the role of $d_{xy}$ orbital discussed above. In short, our results indicate that the ferro-orbital order is unlikely the driving instability of the nematic phase in IBSs. Instead, the instability responsible for the system independent nematic band shift should be the true driver of the nematic phase. It could be an instability with a spin origin or still an orbital origin but in a different form. In any case, our findings provide crucial information on the nematic phase origin issue by eliminating the ferro-orbital order from the candidate list. It should further shed light on the origin of the nematic phase in IBSs.

\noindent
\newline
\textbf{Methods}
\newline
\textbf{Experiments}
\newline
XAS experiments were performed at the beam line 2A of the Pohang Light Source and spectra were recorded in the TEY mode. All spectra were normalized by the incident photon flux intensity measured from a gold mesh and calibrated with respect to $L_{3}$ absorption peak of Fe$_2$O$_3$
 alloy located in front of the analysis chamber. ARPES measurements were carried out at the Beamline 4.0.3 of the Advanced Light Source. Linearly polarized light with the photon energy of 56 eV was used. All crystals were cleaved in situ in a pressure better than 9$\times$10$^{-10}$ Torr for XAS and 4$\times$10$^{-11}$ Torr for ARPES experiments. Samples were detwinned using uniaxial strain which is applied along the tetragonal [110] direction. The best quality data used in the figures were obtained from accidentally detwinned domain.

\noindent
\newline

\begin{addendum}
\item[Acknowledgements]
This work was supported by the Institute for Basic Science in Korea (Grant No. IBS-R009-G2). The work at Pohang University of Science and Technology (POSTECH) was supported by IBS (no. IBS-R014-D1) and the NRF through the SRC (No. 2018R1A5A6075964) and the Max Planck-POSTECH Center (No. 2016K1A4A4A01922028). The Advanced Light Source is supported by the Office of Basic Energy Sciences of the US DOE under Contract No. DE-AC02-05CH11231.

\item[Author Contributions]
C.I.K. and J.S.K. grew the crystals; S.S.H., J.J.S., B.S.K., S.H.C., J.K.J., and S.K. performed ARPES measurements with the support from J.D.D. and W.S.K.; S.S.H., J.J.S., B.S.K., Y.Y.K., performed XAS measurements with the support from Y.H.K.; S.S.H. analyzed the ARPES and XAS data; S.S.H., Y.K.K. and C.K. wrote the paper; Y.K.K. and C.K. are responsible for project direction and planning.

\item[Additional information]
Reprints and permissions information are available online at www.nature.com/reprints. Correspondence and requests for materials should be addressed to Y. K. Kim (yeongkwan@kaist.ac.kr) and C. Kim (changyoung@snu.ac.kr).

\item[Competing financial interests]
The authors declare no competing financial interests.
\end{addendum}

\vspace{10 pt}

\newpage
\begin{figure}
\centerline{\includegraphics[width=0.5\columnwidth,angle=0]{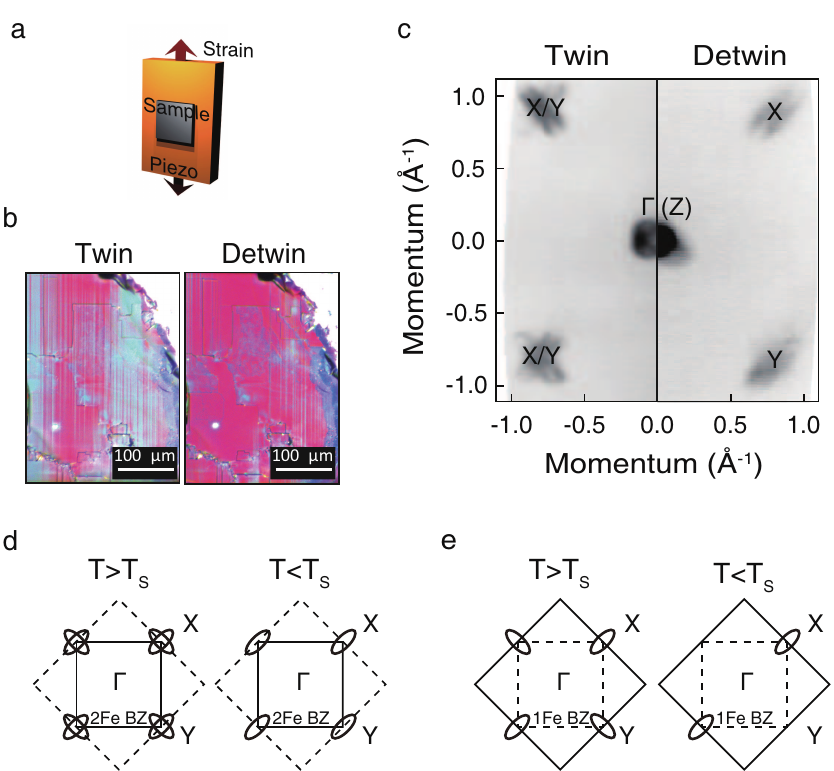}}
\caption{{\bf Electronic structures of pristine and surface electron doped FeSe.} ({\bf a}) Schematic drawing of the piezo sample holder for strain experiments. ({\bf b}) Optical images of twinned and detwinned FeSe single crystal, taken with a polarized microscope. Both images were taken at a temperature below $T_{S}$. ({\bf c}) Corresponding Fermi surface maps. X and Y points are defined by the strain direction of the piezo. Schematic Fermi surface maps in the ({\bf d}) 2Fe BZ and ({\bf e}) 1Fe BZ.}
\label{Fig1}
\end{figure}

\begin{figure*}
\centerline{\includegraphics[width=0.8\columnwidth,angle=0]{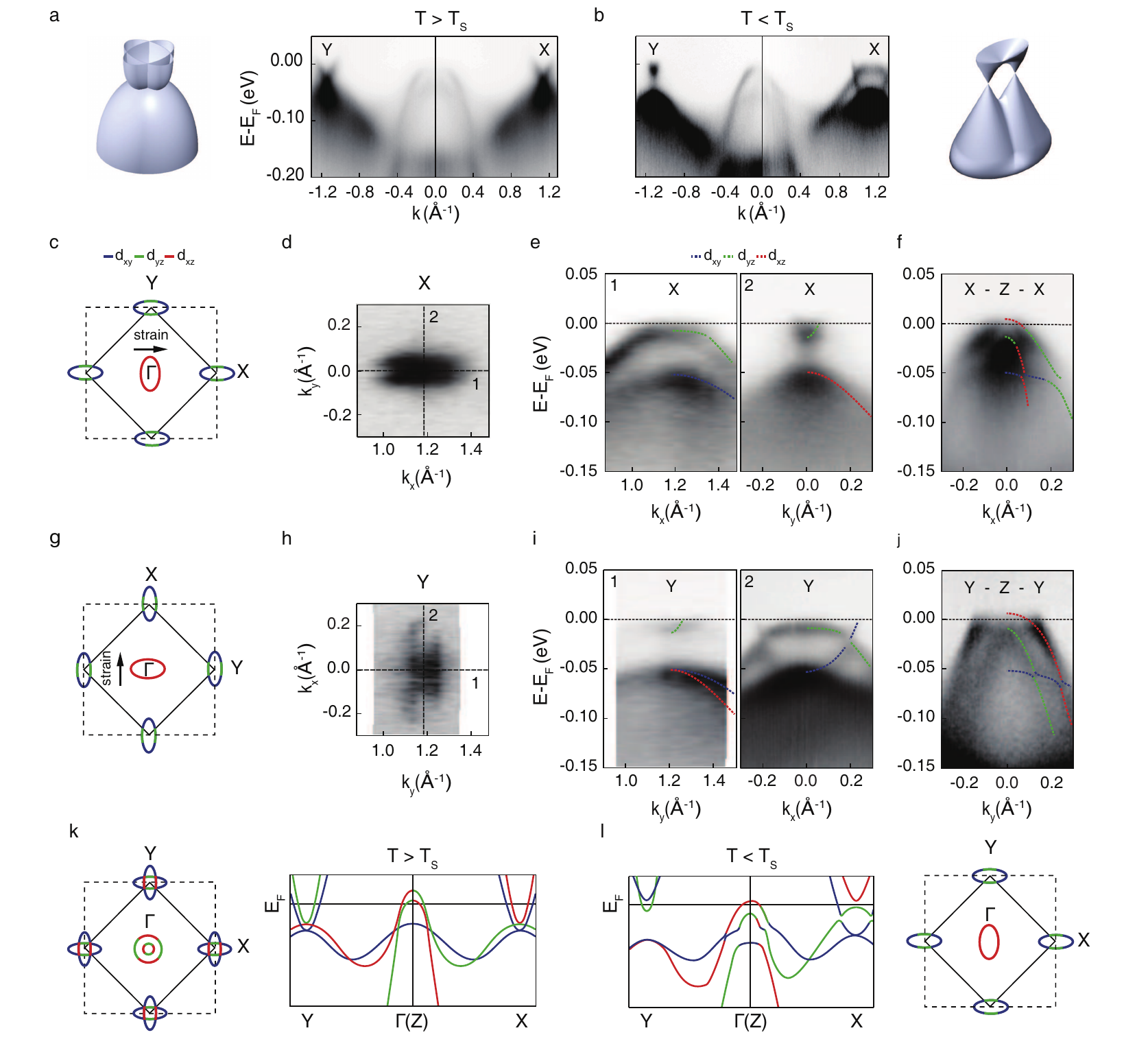}}
\caption{{\bf Dispersions and orbital characters of bands via polarization dependent ARPES.} ({\bf a})-({\bf b}) 3D Schematic band dispersions and high symmetry cuts along the Y-Z-X direction above and below $T_{S}$. All the data were taken with $s$-polarized 56 eV light. ({\bf c}) Schematic Fermi surfaces with orbital characters in the nematic state in the 2Fe BZ scheme. ({\bf d}) Fermi surface map around the X point. ({\bf e}) High symmetry cuts along the $k_{x}$- and $k_{y}$-directions near the X point. The cut directions are shown in ({\bf d}). The overlaid dashed lines are band dispersions with color coded orbital characters. ({\bf f}) High symmetry cut along the X-Z-X direction near the zone center. ({\bf g})-({\bf j}) Similar measurements but with the sample rotated by 90 degree (light polarization along ${a}$-direction). Fermi surface map and high symmetry cuts are now shown for the Y point. ({\bf k})-({\bf l}) Schematic Fermi surfaces and band dispersions with orbital characters along the Y-Z-X direction above and below $T_{S}$.}
\vspace*{-0.5cm}
\end{figure*}

\begin{figure}
\centerline{\includegraphics[width=0.66\columnwidth,angle=0]{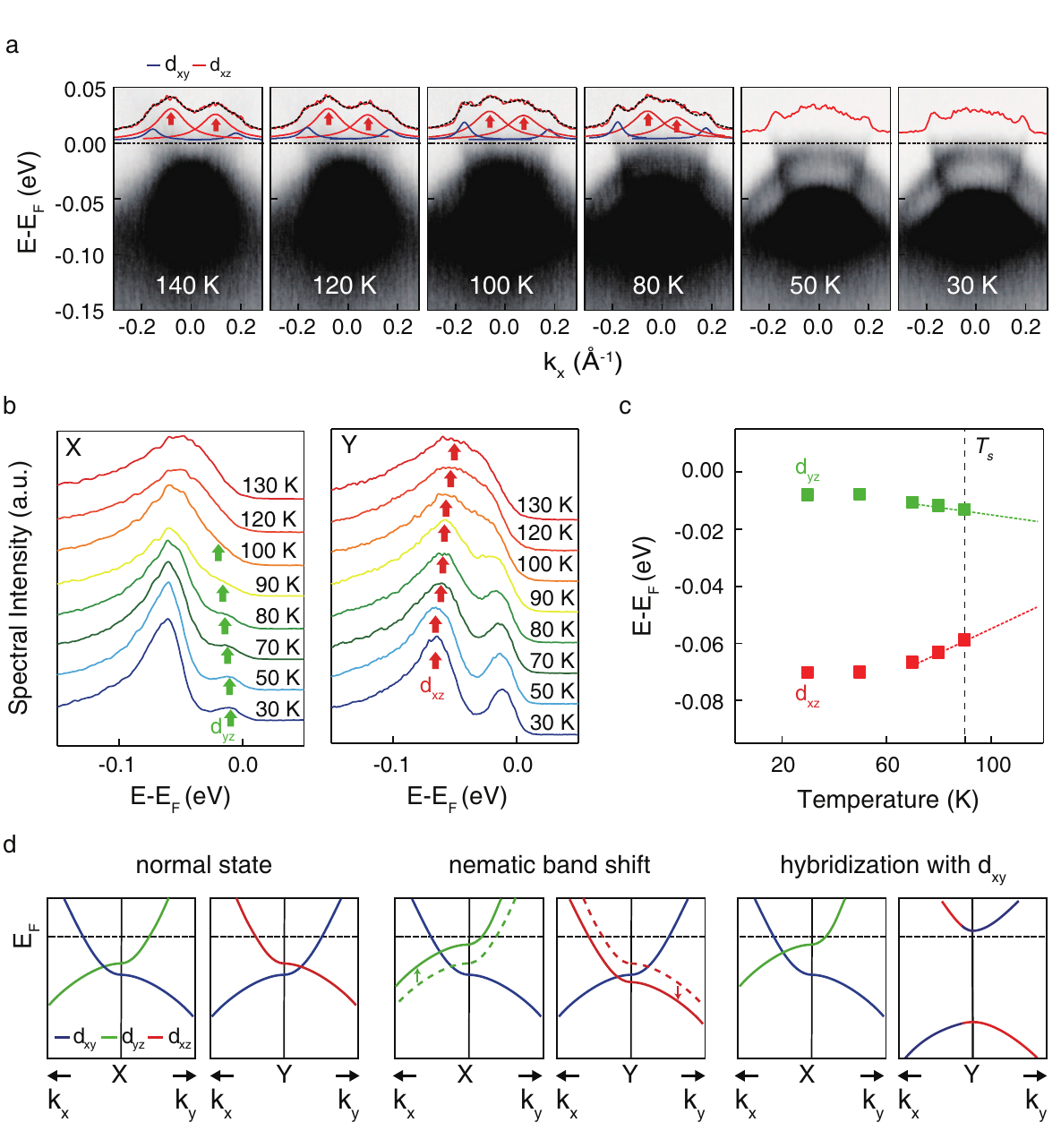}}
\caption{{\bf Temperature evolution of the electronic structure.} ({\bf a}) Temperature dependent ARPES data along the $k_{x}$-direction near the Y point. Momentum distribution curves (MDCs) at 5 meV above $E_{F}$ are also shown at the top of the figure. Underlying $d_{xy}$ and $d_{xz}$ contributions are shown in blue and red curves, respectively. Peak positions of the $d_{xz}$ band are indicated by red arrows. ({\bf b}) Temperature dependent energy distribution curves (EDCs) at X and Y points, showing upward (downward) shift of the $d_{yz}$ ($d_{xz}$) band upon cooling. ({\bf c}) The peak position of the $d_{xz}$ (red square) and $d_{yz}$ (green square) band as a function of temperature. ({\bf d}) Schematic illustration of the band reconstruction at the zone corner across the nematic phase transition.}
\vspace*{-0.5cm}
\end{figure}

\begin{figure}
\centerline{\includegraphics[width=0.45\columnwidth,angle=0]{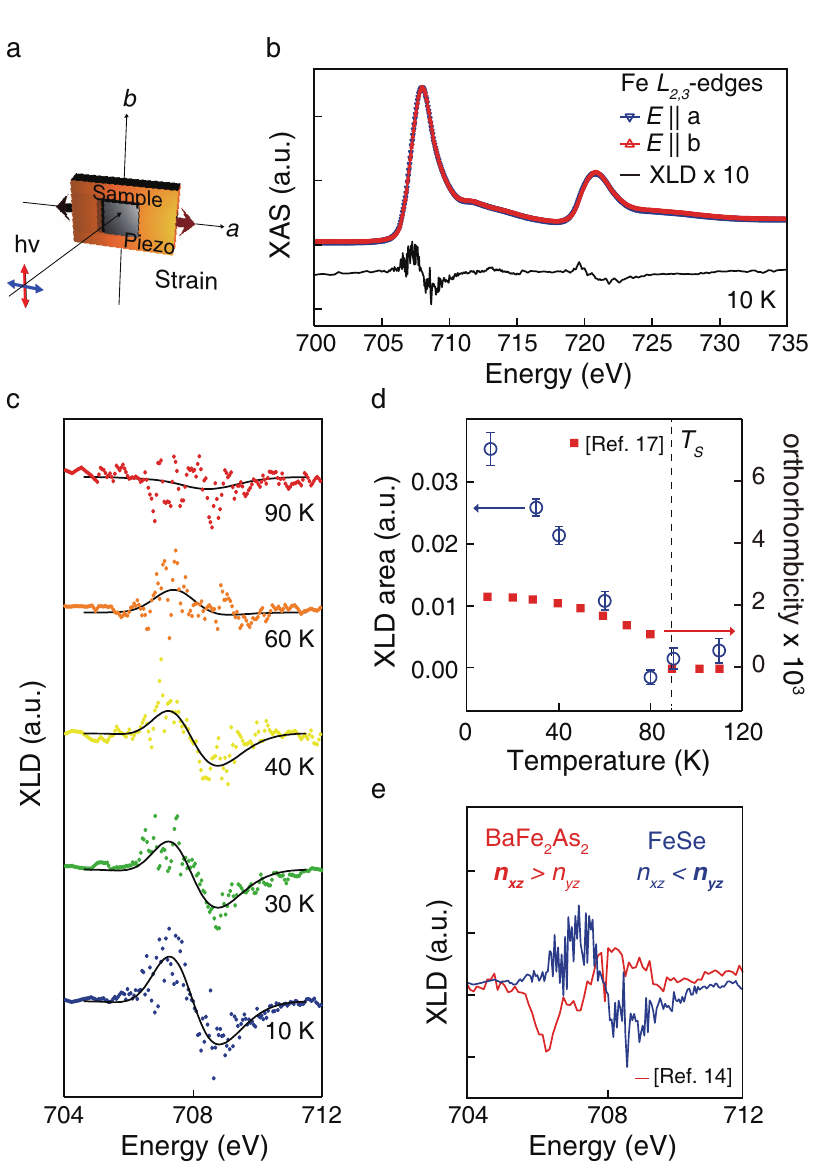}}
\caption{{\bf Observation of orbital occupancy by XLD.} ({\bf a}) Schematic illustration of the experimental geometry. Beam is incident normal to the sample surface. $a$- and $b$-axis are determined by the strain direction of the piezo. ({\bf b}) Fe $L$ edge absorption spectra from detwinned FeSe taken at 10 K with E{$\parallel$}a (blue, inverted triangle) and E{$\parallel$}b (red, triangle) polarizations. The XLD (black curve) is the difference (E{$\parallel$}a - E{$\parallel$}b). XLD spectrum is multiplied by 10 for a better view. ({\bf c}) XLD spectra at various temperatures. Overlaid black solid lines are fitting results with two Gaussian functions. ({\bf d}) Temperature dependence of XLD area (left, blue circle) and orthorhombicity (right, red square). XLD area is calculated by integrating the absolute value of the fit curves in Figure 4 ({\bf c}). The error bars are obtained from the standard fitting error. The orthorhombicity of FeSe is from diffraction measurement result \cite{xrd}. ({\bf e}) XLD spectrum of FeSe (blue) at 10 K and BaFe$_{2}$As$_{2}$ (red, \cite{yk}) at 30 K. The inset above is a schematic figure for energy levels of two materials.}

\end{figure}


\begin{thebibliography}{1}
\item[{\bf References}]

\bibitem{nem} Fernandes, R. M., Chubukov, A. V. \& Schmalian, J. What drives nematic order in iron-based superconductors?  \emph{Nat. Phys.} {\bf 10}, 97-104 (2014).

\bibitem{cup} Daou, R. \emph{et al}. Broken rotational symmetry in the pseudogap phase of a high T$_c$ superconductor. \emph{Nature} {\bf 463}, 519-522 (2010).

\bibitem{ibsnem} Kasahara, S. \emph{et al}. Electronic nematicity above the structural and superconducting transition in BaFe$_2$(As$_{1-x}$P$_{x}$)$_2$. \emph{Nature} {\bf 486}, 382-385 (2012).

\bibitem{nomag1} Medvedev, S. \emph{et al}. Electronic and magnetic phase diagram of $\beta$ Fe$_{1.01}$ with superconductivity at 36.7 K under pressure. \emph{Nat. Mater} {\bf 8}, 630-663 (2009).

\bibitem{nomag2} Baek, S. H. \emph{et al}. Orbital-driven nematicity in FeSe. \emph{Nat. Mater} {\bf 14}, 210-214 (2015).

\bibitem{fsres} Tanatar, M. A. \emph{et al}. Origin of the Resistivity Anisotropy in the Nematic Phase of FeSe. \emph{Phys. Rev. Lett.} {\bf 117}, 127001 (2016).

\bibitem{stm1} Sprau, P. O. \emph{et al}. Discovery of orbital-selective Cooper pairing in FeSe. \emph{Science} {\bf 357}, 75-80 (2017).

\bibitem{stm2} Kostin, A.  \emph{et al}. Imaging orbital-selective quasiparticles in the Hund’s metal state of FeSe. \emph{Nat. Mater} {\bf 17}, 896-874 (2018).

\bibitem{ferro} Shimojima, T. \emph{et al}. Lifting of $xz$/$yz$ orbital degeneracy at the structural transition in detwinned FeSe. \emph{Phys. Rev. B.} {\bf 90}, 121111 (2014).

\bibitem{dwave} Zhang, P. \emph{et al}. Observation of two distinct $d_{xz}$ /$d_{yz}$ band splittings in FeSe. \emph{Phys. Rev. B.} {\bf 91}, 214503 (2016).

\bibitem{uni} Watson, M. D. \emph{et al}. Evidence for unidirectional nematic bond ordering in FeSe. \emph{Phys. Rev. B.} {\bf 94}, 201107 (2016).

\bibitem{sr} Suzuki, Y. \emph{et al}. Momentum-dependent sign inversion of orbital order in superconducting FeSe. \emph{Phys. Rev. B.} {\bf 92}, 205117 (2015).

\bibitem{det} Watson, M. D. \emph{et al}. Electronic anisotropies revealed by detwinned angle-resolved photoemission spectroscopy measurements of FeSe. \emph{New J. Phys.} {\bf 19}, 103021 (2017).

\bibitem{yk} Kim, Y. K. \emph{et al}. Existence of orbital order and its fluctuation in superconducting  Ba(Fe$_{1-x}$Co$_{x}$)$_{2}$As$_{2}$ single crystals revealed by x-ray absorption spectroscopy. \emph{Phys. Rev. Lett.} {\bf 111}, 217001 (2013).

\bibitem{ccl} Lee, C. -C., Yin, W. -G. \&  Ku, W. Ferro-Orbital Order and Strong Magnetic Anisotropy in the Parent Compounds of Iron-Pnictide Superconductors. \emph{Phys. Rev. Lett.} {\bf 103}, 267001 (2009).

\bibitem{ccc} Chen, C.-C. \emph{et al}.Orbital order and spontaneous orthorhombicity in iron pnictides. \emph{Phys. Rev. B.} {\bf 82}, 100504 (2010).

\bibitem{xrd} Margadonna, S. \emph{et al}. Crystal structure of the new FeSe$_{1-x}$ superconductor. \emph{Chem. Commun.} {\bf 2008}, 5607-5609 (2008).

\bibitem{weak1} Mazin, I. I., Singh, D. J., Johannes, M. D. \& Du, M. H. Unconventional superconductivity with a sign reversal in the order parameter of LaFeAsO$_{1-x}$F$_x$. \emph{Phys. Rev. Lett.} {\bf 101}, 057003 (2008).

\bibitem{weak2} Dong, J.  \emph{et al}. Competing orders and spin-density-wave instability in La(O$_{1-x}$F$_x$)FeAs. \emph{Euro. Phys. Lett.} {\bf 83}, 27006 (2008).

\bibitem{strong1} Yin, Z. P. \& Pickett, W. E. \emph{et al}.Crystal symmetry and magnetic order in iron pnictides: A tight-binding Wannier function analysis. \emph{Phys. Rev. B.} {\bf 81}, 174534 (2010).

\bibitem{strong2} Oh, H., Moon, J., Shin, D., Moon, C.-Y. \& Choi, H. J. Prog. Brief review on iron-based superconductors: are there clues for unconventional superconductivity? \emph{Prog. Supercond.} {\bf 13}, 65-84 (2011).

\bibitem{resoo1} LV, W. \& Phillips, P. \emph{et al}. Orbitally and magnetically induced anisotropy in iron-based superconductors. \emph{Phys. Rev. B.} {\bf 84}, 174512 (2011).

\bibitem{resoo2} Liang, S., Alvarez, G., Şen, C., Moreo, A.  \& Dagotto, E. \emph{et al}. Anisotropy of Electrical Transport in Pnictide Superconductors Studied Using Monte Carlo Simulations of the Spin-Fermion Model. \emph{Phys. Rev. Lett.} {\bf 109}, 047001 (2012).




\end{thebibliography}
\end{document}